# Deep Learning for Asset Bubbles Detection[*]

Oksana Bashchenko[†]and Alexis Marchal[‡]

February 11, 2020

We develop a methodology for detecting asset bubbles using a neural network. We rely on the theory of local martingales in continuous-time and use a deep network to estimate the diffusion coefficient of the price process more accurately than the current estimator, obtaining an improved detection of bubbles. We show the outperformance of our algorithm over the existing statistical method in a laboratory created with simulated data. We then apply the network classification to real data and build a zero net exposure trading strategy that exploits the risky arbitrage emanating from the presence of bubbles in the US equity market from 2006 to 2008. The profitability of the strategy provides an estimation of the economical magnitude of bubbles as well as support for the theoretical assumptions relied on.

## 1. Introduction

Asset bubbles have preoccupied investors nearly since the beginning of financial markets. From the Tulip mania of 1636-1637 all the way to the more recent financial depression of 2008, including many others in between. Recently, bubble concerns have resurfaced with the rise of crypto assets experiencing rapid increase and violent crashes in prices. The concept of a bubble (defined as a deviation of the price from the fundamental value) might be well known for market participants but its detection remains a challenging task because the fundamental value is not observable.

The fundamental value is the expected sum of future cash-flows or benefits offered by the security, appropriately discounted for time and risk. The difficulty comes from the fact that one always needs some assumptions in order to make a statement about it. One approach is to build a theoretical model for the fundamental value but then the bubble detection procedure would suffer from a joint-hypothesis problem (you are simultaneously testing the presence of a bubble and the assumptions of the model). A second approach is to use almost model-free techniques from mathematical finance under the assumption of a continuous-time economy. Essentially, this boils down to classifying stock prices as being either true or strict local martingales. However to do so it is necessary to estimate the volatility function of a diffusion process given discrete time-series data and this method suffers from estimation errors.

This paper aims at providing a bubble detection methodology that outperforms the existing one under a continuous-time paradigm. To the best of our knowledge, we are the first to provide a certain number of findings to the literature. Our first contribution is to show that long short-term memory (LSTM) networks outperform current methods at detecting asset bubbles both in terms of accuracy and computational time. The second contribution is to provide an evaluation of the method proposed in R. Jarrow, Kchia, and Protter (2011) and Obayashi, Protter, and Yang (2017) in a controlled environment. Indeed, by using simulated data we know exactly when an asset bubble occurs and are able to assess the accuracy of their method (and in a second step compare it to ours). Finally, our last contribution is to assess the economic significance of these bubbles. We essentially detect them on real data and build a long-short trading strategy to assess how much a trader could potentially earn by exploiting this risky arbitrage.

The profitability of the strategy provides support for theoretical foundations that use strict local martingales to detect bubbles. It empirically shows that it is valid to consider the price as a continuous-time process that can only be observed at discrete random times (the trading times), as postulated in R. Jarrow and Protter (2012).

The rest of the paper is organized as follows. Section 2 describes the literature. Section 3 lays out the theoretical foundations for characterizing bubbles in continuous-time and presents the existing methods for detecting them. Section 4 explains the main idea behind the architecture and the training of the network. Section 5 assesses the performance of our method on out-of-sample simulated data. Section 6 applies the algorithm to detect bubbles using real data and builds a trading strategy that exploits this risky arbitrage. Section 7 discusses the implications of a discrete-time paradigm for our results. Finally section 8 concludes.

## 2. LITERATURE REVIEW

With the focus being mostly on the martingale theory of bubbles, our paper is largely inspired by the insights of R. A. Jarrow, Protter, and Shimbo (2007) and R. A. Jarrow, Protter, and Shimbo (2010). In these studies the authors provide conditions for bubbles to exist in complete and incomplete markets respectively. Importantly, they show that a non-trivial bubble can be of three types. Two of them can exist only in infinite horizon economies, thus being less relevant for empirical applications. The third type of bubbles, that is the only type viable in the finite horizon framework, is proven to be represented as a strict local martingale under a risk-neutral measure. We are interested in detecting bubbles of that third type in data, and it explains our inclination to work with a continuous-time framework. The interested reader may find more details about the mathematical foundation of bubbles in continuous-time in the work of Protter (2013), while a general review of the asset price bubbles is proposed in R. A. Jarrow (2015).

For diffusion processes with volatility being a function of the price solely, Delbaen and Shirakawa (2002) and Mijatović and Urusov (2012) provide a theoretical test based on the instantaneous volatility to distinguish between a true and a strict local martingale. The difficulty in applying this test for data is that the functional form of the volatility has to be known. R. Jarrow, Kchia, and Protter (2011) try to solve this problem by proposing one parametric and multiple non-parametric methods for estimating it. One limitation of their paper is that the functional form of the volatility has to stay the same over the whole sample period. The recent work of Obayashi, Protter, and Yang (2017) mitigates this constraint by allowing the functional form to vary among different time periods. This results into adding more flexibility to the method, but deteriorating the quality of each period estimation, leaving space for further improvement. We propose a new direction for this literature, departing from the classical statistical test by using deep learning techniques.

Machine learning methods are gaining a huge popularity in financial research, both among academicians and practitioners. With financial data being mostly represented as time-series, the family of recurrent neural networks is the logical choice for the majority of problems in the field. A prominent member of this class is the long short-term memory network (LSTM) proposed by Hochreiter and Schmidhuber (1997). It is able to remember important information that was faced in the far past and forget unimportant recent one. This type of network is so far the most popular for tasks involving complex historical patterns of data.

## 3. BUBBLES IN CONTINUOUS-TIME

There are two paradigms on modeling asset prices: discrete or continuous-time. Often, both frameworks are equivalent in the sense that they describe the same phenomenon. For instance, for derivative pricing a discrete-time binomial model approximates the Black-Scholes formula which relies on a continuous-time framework. However this equivalence is in general not true.

Whether we adopt a discrete or continuous-time approach greatly changes the theoretical framework and assumptions required for bubbles detection. In a discrete-time economy, bubbles can only exist if the time horizon is infinite. Moreover, the only way to detect them is by relying on a model for the unobserved fundamental value, creating a joint-hypothesis problem. In continuous-time, bubbles can exist even in finite horizon economies and we only need to assume a certain stochastic differential equation (SDE) describing the evolution of the price. We can assess whether the SDE accurately describes the price changes independently from testing for the presence of bubbles (hence eliminating the joint-hypothesis problem). In both cases, we are forced to make a relatively strong assumption: either infinite time horizon or continuous-time. In this paper we have decided to make the latter assumption. This implies that we use objects (strict local martingales) which only exist in continuous-time models but not in their discrete-time counterparts.

It is therefore natural to wonder: how valid is this assumption? And how relevant are the results for the real world? In reality it is true that, despite the fact that high-frequency firms trade with incredibly short intervals, it still happens at discrete times. However, a trade can happen at any point on the half-line $[0,\infty)$ that represents time. At the tick level, the time between trades is not uniformly spaced as pointed out by R. Jarrow and Protter (2012). This contrasts with a discrete-time model where the time grid is fixed and trading is forced to happen at uniformly spaced dates. This is why it is more plausible to imagine that the price process evolves in continuous-time and we are only able to observe its values at random stopping times (when a trade takes place). Moreover, even though our objects of interest only exist in continuous-time, they can be approximated by discrete-time processes (see appendix B.3 for more details). A longer discussion on this topic is available in Protter (2013) in section 11.

Nevertheless, if the reader is a firm believer of the discrete-time framework, we can still give meaning to our results. We show that what we detect are true martingales but with significantly skewed distribution. We discuss in depth the link with a discrete-time economy in section 7.

### 3.1. THEORETICAL FRAMEWORK

We essentially use the set-up described in the annual review of R. A. Jarrow (2015) which is also common to most of the papers in this literature. We keep the theoretical framework to a minimum to obtain a concise paper but we still ensure that the reader can understand our main contributions.

Let $(\Omega, \mathcal{F}, \mathbb{F}, \mathbb{P})$ be a filtered probability space where the filtration satisfies the usual hypothesis (Protter (2001)). Time is continuous and the economy lives over a compact time interval $[0, T]$ with $T < \infty$. As discussed earlier, we only study bubbles that occur in finite horizon economies. There exist two assets. The risky asset price is denoted by $S_t$ and its real world dynamics is described by

$$dS_t = b(t, S_t)dt + \sigma(t, S_t)dB_t, \quad S_0 = s_0 \tag{3.1}$$

where $B_t$ is a one-dimensional $\mathbb{P}$-Brownian motion. We denote by $\tau \leq T$ the random terminal date[1] of this asset. The second asset is locally riskless and pays the risk-free rate $r_t$. Its value is given by the solution of the following differential equation

---

[1]This date can be thought as a bankruptcy or when the shareholders decide to liquidate the assets of the firm for instance.

$$dS_{0t} = S_{0t} r_t dt, \quad S_{00} = 1. \tag{3.2}$$

The discounted price of the risky asset is defined in the classical way as

$$\hat{S}_t \triangleq \frac{S_t}{S_{0t}}. \tag{3.3}$$

As mentioned before, an asset is in a bubble when the spot price is trading above the fundamental value. So first we need to clarify what the fundamental value is. We assume that the cumulative dividends process $D = (D_t)_{0 \leq t \leq \tau}$ is an adapted càdlàg non-negative semimartingale. Then the fundamental value $S_t^\star$ is defined in a standard way as the expected sum of discounted cash-flows to be paid. Let $\mathcal{M}$ be a non-empty set of equivalent local martingale measures (ELMM). We fix[2] a measure $\mathbb{Q} \in \mathcal{M}$ and write

$$\hat{S}_t^\star \triangleq \mathbb{E}_t^{\mathbb{Q}} \left[ \int_t^\tau \frac{dD_s}{S_{0s}} + \hat{L}_\tau \right] \tag{3.4}$$

where $\hat{L}_\tau \geq 0$ is the discounted liquidation value of the asset at the terminal date. We formally define a bubble process as

$$\beta_t = S_t - S_t^\star. \tag{3.5}$$

We say that an asset experiences a bubble whenever $\beta_t > 0$. The next theorem characterizes this phenomenon. For simplicity of exposition, we assume no dividends from now on.

**Theorem 3.1** (Obayashi, Protter, and Yang (2017)). *A risky asset price process S is undergoing bubble pricing on the compact time interval $[0, T]$ if and only if under the chosen risk neutral measure the discounted bubble process $\hat{\beta}$ is not a martingale but is a strict local martingale. This is equivalent to the discounted price process $\hat{S}$ being a strict local martingale (since $\hat{S}^\star$ is always a martingale) and not a martingale.*

Theorem 3.1 provides us with a simple criterion for detecting bubbles by identifying strict local martingales. Now we can fully understand the implication of our assumptions since we aim at detecting strict local martingales which only exist in continuous-time. A benefit of this setting is that we do not need to model the fundamental value, all we need to know is whether the discounted stock price is a strict local martingale or not.

Under $\mathbb{Q}$ the process $\hat{S}_t$ has the following dynamics

$$d\hat{S}_t = \sigma(t, S_t) S_{0t}^{-1} dB_t^{\mathbb{Q}}. \tag{3.6}$$

---

[2] For concerns regarding the incompletness of the market we refer to the discussion in R. A. Jarrow (2015).

By Girsanov theorem, the diffusion coefficient stays the same under any equivalent change of measure, so without loss of generality we can in fact work with any $\mathbb{Q}$. Now that we have the risk-neutral dynamics, we need to know how to differentiate between true and strict local martingales. From now on, we assume the interest rate to be constant and equal to zero for simplicity of exposition. This will not impact our results on simulated data. Regarding the application on real data where rates are not constant, analogous results will hold under some technical conditions. A discussion can be found in Protter (2013).

Thanks to the work of Delbaen and Shirakawa (2002) we have a simple condition to distinguish between true and strict local martingales. For a process of the form

$$dX_t = b(X_t)dB_t^{\mathbb{Q}}, \tag{3.7}$$

we first define an integral

$$I(\varepsilon) \triangleq \int_\varepsilon^\infty \frac{x}{b(x)^2} dx. \tag{3.8}$$

Then depending on whether this integral converges or not two cases can arise:

- $X$ is a strict local martingale (i.e. experiences a bubble) if and only if $I(\varepsilon) < \infty$ for all $\varepsilon > 0$.

- $X$ is a true martingale (i.e. does not experience a bubble) if and only if $I(\varepsilon) = \infty$ for all $\varepsilon > 0$.

At this point, we need to make assumptions about the diffusion coefficient $\sigma(t,s)$ in the SDE (3.6). The bubble testing procedure requires us to obtain this coefficient solely as a function of $s$, and not as a function of time. However assuming that the volatility function only depends on the stock price and that it never changes through time is unrealistic. This would imply for instance that if a stock is in a bubble, it will stay in this state forever. We therefore follow Obayashi, Protter, and Yang (2017) and assume a regime change model for the diffusion coefficient such that

$$\sigma(t,x) = \sigma_1(x)\mathbb{1}_{[t_1,t_2]}(t) + \sum_{i=2}^n \sigma_i(x)\mathbb{1}_{(t_i,t_{i+1}]}(t). \tag{3.9}$$

Note that time intervals do not need to be equally spaced. Now over a given time interval $(t_i, t_{i+1}]$ we will be able to estimate the diffusion coefficient as a function of $s$ only and perform the integral test locally using (3.8). In a theoretical model, the computation of $I(\varepsilon)$ is (easily) done. However, applying this test on real data is complicated by the fact that $\sigma(x)$ is not known. Its recovery becomes the cornerstone of the analysis. The tricky point of this estimation is that the functional form may vary over the ex-ante unknown time intervals. Another difficulty is that only data on a bounded interval is available for this estimation, while what matters for $I(\varepsilon)$ is the behavior of $\sigma(x)$ when $x \to \infty$.

For the readers familiar with the estimation of historical volatility, it is important to emphasize that the task is *not* to estimate the volatility through time but to estimate the functional form of volatility $\sigma(x)$ with respect to the price $x$ over some time interval. This implies that volatility estimators like realized variance (RV) are of no use for this task.

As another side note, it is worth noticing that $\beta_t \geq 0$ for any $t$ as shown in R. A. Jarrow (2015) in theorem 3 for instance. This means that in this framework, the stock price might be above the fundamental value but never below. It implies that when later we build a trading strategy using real data, we will always short stocks that are classified as being in a bubble.

### 3.2. CLASSICAL STATISTICAL METHOD FOR BUBBLE DETECTION

In this section, our goal is to motivate the necessity of a neural network to detect bubbles by showing that it significantly improves the detection rate. We do so by constructing a simple example on simulated data where the stock volatility is assumed to only have two regimes. We show that already here the method described in the literature often underperforms. This example is unrealistically simple but sufficient to prove our point. Later we will use a more complex structure for the data-generating process when making the final comparison of both methods.

We now discuss the method presented in Obayashi, Protter, and Yang (2017). The authors decide to estimate $\sigma(x)$ over a rolling window of 21 days and propose two estimators. The first one being a parametric estimator and a second non-parametric one based on the local time of a Brownian motion. In that case, they can analyze when a bubble is born and when it bursts (and ultimately construct a distribution of the lifetime of a bubble).

In our paper we will only present and benchmark our network to the parametric estimator (PE). Given the fact that we will assess the performance of the two methods on simulated data, we always know the true data-generating process. We use the same family of functions for data simulations and for the parametric estimator. Since the PE uses a function from the true family, its performance will be higher than a non-parametric estimator. The latter might lead to better performance on real data where the functional family is unknown. However, with real data we cannot compare the methods because we do not know the true volatility regime.

In order to simulate data, we first assume a functional form for $\sigma(x)$. We choose the well-known power family[3]

$$\sigma(x) = \gamma_0 x^{\gamma_1} \tag{3.10}$$

where $\boldsymbol{\gamma} = (\gamma_0, \gamma_1)$ is constant over a given time interval $(t_i, t_{i+1}]$ but is allowed to change from one time interval to another. Given this function, the stock price $S_t$ is a true martingale only for $1/2 < \gamma_1 \leq 1$. For $\gamma_1 > 1$ it is a strict local martingale. In this simplified example we simulate price paths by using only two regimes:

**R1** When the stock is not in a bubble (i.e is a true martingale) and so we set $\gamma_1 = 0.9$.

---

[3]This functional form implies that we model the stock price with a constant elasticity of variance (CEV) model.

**R2** When the stock is in a bubble (i.e is a strict local martingale) and we set $\gamma_1 = 1.1$.

A Markov chain decides on the random times when the stock transits back and forth between regimes **R1** and **R2**. It is clear that the scaling parameter $\gamma_0$ does not impact the convergence of the integral in (3.8) so we set $\gamma_0 = 0.15$ for both regimes for simplicity. Once the data is simulated, we need to estimate $\boldsymbol{\gamma}$.

The above-mentioned paper relies on a parametric estimator originally introduced in Genon-Catalot and Jacod (1993) which is given by

$$\hat{\boldsymbol{\gamma}}^{\text{PE}} = \operatorname{argmin} \frac{1}{n} \sum_{i=1}^{n} \left( \gamma_0^2 S_{t_{i-1}}^{2\gamma_1} - \frac{n}{T} \left( S_{t_i} - S_{t_{i-1}} \right)^2 \right)^2. \tag{3.11}$$

Since the data is simulated we know exactly when the stock was in a bubble and we compare this information to the one given by the estimators. To do so, we construct the following random variable

$$\xi_t \triangleq \mathbb{1}_{\{1/2 \leq \gamma_{1,t} \leq 1\}} \tag{3.12}$$

which takes the value one when the stock is a true martingale and zero otherwise. Then we denote by $\hat{\xi}_t^{\text{PE}}$ and $\hat{\xi}_t^{\text{NN}}$ the parametric estimator[4] (PE) and the neural network estimator (NN) respectively.

The results for a single price path are displayed in figure 3.1 which contains four subplots. The first one simply shows the log-returns of the price through time during the different market regimes. The second subplot displays $\xi$ which takes the value 0 when the stock experiences a bubble and otherwise is equal to 1. Here we see that over the whole period, the stock experienced 3 bubbly regimes and 3 "normal" regimes. Then the last two subplots display the estimation of $\xi$ according to the parametric estimator available in the literature and the neural network respectively. A direct visual inspection reveals that the neural network largely outperforms at detecting bubbles (i.e $\hat{\xi}_t^{\text{NN}}$ is a better estimator of $\xi$).

---

[4] In order to obtain $\hat{\xi}_t^{\text{PE}}$ we follow the methodology described in Obayashi, Protter, and Yang (2017). More precisely, we first use the estimator described in equation (3.11) and then apply a Hidden Markov Model to smooth the signal and use the probability matrices from their paper.

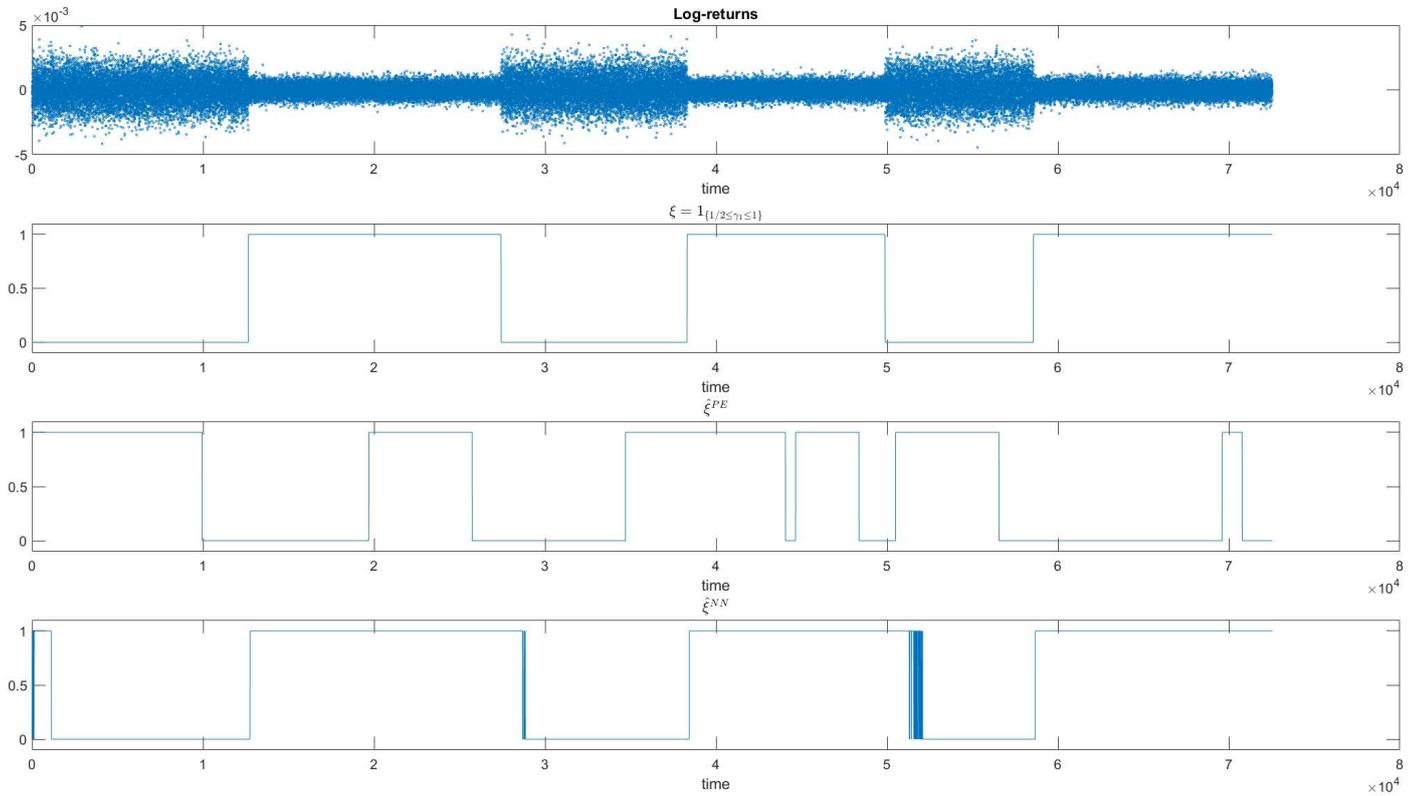

Figure 3.1: Simulated data showing the drawback of the parametric estimator (PE)

Why is it the case? If $S_t$ stays in one specific regime (either true martingale or strict local martingale) over the whole sample path, then the statistical estimator performs quite well. However the statistical method starts to be challenged as soon as it has to perform the estimation of $\sigma(x)$ during a period when the regime changed. The reason is very intuitive. When using a rolling window around a regime change, the estimator will be using data points generated during the old regime as well as data points belonging to the new regime. This implies that it will fail to correctly estimate $(\gamma_0, \gamma_1)$ around these times. Choosing a large window will lead to more stable coefficients however it will fail to identify the exact time of a regime change (or even miss some short lived regime changes). A small window leads to unstable estimations. Obayashi, Protter, and Yang (2017) mitigate this problem by combining their estimator with a Hidden Markov Model used to smooth the bubble signal. Nevertheless, as we see here it still underperforms a neural network.

How does the network manage to overcome this problem? Intuitively, it is first able to detect the times of regime changes before estimating the parameters $(\gamma_0, \gamma_1)$. By doing so, it does not use data from another regime. In a given identified regime, it includes all the data points from this regime for the estimation, using potentially more appropriate data than with a rolling window and thus improving the estimation.

This simple example is easy to understand but is far from reality. For instance the time-series of returns in figure 3.1 does not resemble real returns. This is why a thorough comparison of both methods in realistic conditions will be done in section 5 after introducing the neural network and the training process. However already in this toy example the statistical method struggles. In a more sophisticated environment its performance will deteriorate even further.

## 4. NETWORK ARCHITECTURE AND TRAINING

The long short-term memory network we choose for bubble classification is a special type of recurrent neural network that was introduced in Hochreiter and Schmidhuber (1997). By construction, LSTM networks are well suited for remembering long-term dependencies and thus for handling time-series data. We use a standard LSTM, motivated by the main finding of Greff et al. (2017). They present the largest comparison study for different modifications of LSTM networks and show that improvements over the plain vanilla LSTM are marginal. Our network consists of six layers: an input layer, followed by two bidirectional LSTM layers with 100 hidden neurons each, a fully connected layer, a softmax layer, and a classification output layer. An interested reader may refer to appendix A to find a brief intuition about neural networks in general and LSTM in particular, as well as a detailed layers description.

Now that the structure of the network is chosen, it has to be trained to detect strict local martingales. Our methodology is conceptually simple. We first generate training data and since it is simulated, we know exactly when the stock was a true martingale (TM) or a strict local martingale (SLM). Then we label each regime as "TM" or "SLM" and train the network on them. Finally the network is ready to classify new time-series.

To create the training data, we simulate paths of the price process (3.6) by discretizing the stochastic differential equation governing it. The training data set as a whole can be composed by multiple time-series possibly generated from different processes by assuming different functional forms for $\sigma(x)$. Using a variety of them allows the network to concentrate only on the specific feature (TM or SLM) of the data point, preventing over-fitting. Another benefit of this method is that we can generate virtually an unlimited quantity of labelled data to train on, enhancing the network performance at no cost.

In order to incorporate stylized facts of equity data, we have decided to model $\sigma(t, x)$ from (3.9) with a Markov chain. Locally, every $\sigma_i(x)$ is modeled using the power function (3.10). To each state of the chain corresponds a different value for $\boldsymbol{\gamma}$. For half of the states we impose $\gamma_1 \leq 1$ while for the other half we set $\gamma_1 > 1$. The Markov chain will control when to switch from $\sigma_i(x)$ to $\sigma_{i+1}(x)$. The transition matrix is made time-varying to prevent the network from learning the probabilities instead of estimating $\sigma(x)$.

Neural networks are considered by some people as a black box. One of our aims is to make the usage of a network as transparent as possible. In section 6 we build a trading strategy and try to predict future price changes. However we do not simply feed past price data to a network and ask for a prediction. By doing so, we would not know what factors does the network use in order to make predictions. This approach would be totally model-free but would also be a sort of black box. Instead, we impose some structure on what the network learns by combining it with a model for the stock price. First we assume that the price is governed by a stochastic differential equation. We impose that the network recognizes certain functional forms for the diffusion coefficient (see eq. (3.10) for instance) and we link these functions to the mathematical theory of asset bubbles. Given the prediction of the theory, we can build a trading strategy.

## 5. Performance on simulated data

In this section we assess the performance of the neural network using out-of-sample simulated data. We start by generating new time-series, using parameters[5] that the network was never trained on. Then we classify the data by using the parametric estimator (3.11) and the neural network. Both methods were tested using 150 Monte-Carlo paths, each individual path consisting of data generated at a 2 minutes frequency for 3 years. The results are displayed in table 5.1. The row "% detection" refers to the percentage of data points that were correctly classified. Those include classifying the generating process as being a TM when it was a TM and classifying the process as being an SLM when it was an SLM. "% spurious" refers to the percentage of points when a method was mistaken. Those include classifying the process as being a TM when in reality it was an SLM and classifying the process as being an SLM when in fact it was a TM.

|             | Network | Obayashi, Protter, and Yang (2017) |
|-------------|---------|-------------------------------------|
| % detection | 83.64   | 49.71                               |
| % spurious  | 16.36   | 50.29                               |

Table 5.1: Comparison of the accuracy of the neural network and the statistical estimator from the literature for bubbles detection.

The results show that the neural network outperforms the existing statistical estimator by displaying a higher rate of correct and a lower rate of spurious detections. We have tried multiple different network architectures and tuned the hyperparameters so that to maximize the performance on out-of-sample simulated data. Gu, Kelly, and Xiu (2019) find that more shallow networks outperform deeper ones at forecasting financial time-series due to the low signal-to-noise ratio. We obtain a similar result although we use a network for a classification problem.

At this point, it is worth emphasizing that once the network is trained, it is much faster at classifying data. The parametric estimation (since it relies on an optimization) took approximately 43 hours to accomplish the required task while the network did it in less than one hour.

---

[5]The new parameters are the values for $(\gamma_0, \gamma_1)$ as well as transitions probabilities for the Markov chain governing the volatility regime changes.

# 6. APPLICATION TO REAL DATA

In this section we perform bubble classification on real data using our neural network. The dataset consists of 30 individual stocks included in the Dow Jones index (plus the index itself) over a 3 years period (from 2006 to 2008). We select a sampling frequency of 2 minutes. This is high enough to have sufficiently many data points to precisely estimate the volatility function but yet low enough so that the data is not contaminated with microstructure noise. Here clearly we cannot exactly assess the performance of our test comparatively to the benchmark since the true data-generating process is unknown.

It is important to emphasize that up to now we only considered the dynamics of the discounted stock price $\hat{S}_t$ under the risk-neutral measure. By definition of the measure, this process has no drift. We did that because to determine if a stock is in a bubble all we need is the functional form of the diffusion coefficient in order to compute the integral (3.8). The theory tells us that when the stock price is in a bubble, its discounted value is a strict local martingale under $\mathbb{Q}$. Because it is bounded from below by 0, the discounted price is a $\mathbb{Q}$-supermartingale. This means that $\hat{S}_t$ is expected to decrease (under $\mathbb{Q}$). However, when trading stocks in the physical world, the processes describing their evolution likely have a non-zero drift under $\mathbb{P}$ (see equation (3.1) for the real world dynamics of the stock). Therefore to profit from the existence of bubbles, it is necessary to go long and short in different assets in order to cancel any drift effect.

This leads us to a natural way to test our detection method and assess the economic magnitude of asset bubbles in the equity market. For that we build a risky arbitrage trading strategy that shorts an individual stock when in a bubble and at the same time goes long in an asset that replicates its fundamental value. Therefore the short-leg at any time $t$ of our strategy consists of all the individual stocks that are experiencing a bubble at $t$. The long-leg is the Dow Jones index (which is a proxy for the fundamental value of the short-leg). We invest the same dollar amount in the long and short legs so that we obtain a zero investment strategy. Given that our data covers the beginning of the financial crisis of 2008, the long-short strategy allows us to shield our portfolio from any market-wide shock (we have a zero net directional exposure).

Before proceeding to the results, it is important to mention that real data are probably not generated by a process as simple as (3.1). There is substantial evidence that individual stocks experience frequent jumps (see for instance Andersen, Bollerslev, and Dobrev (2007)) and are therefore best modeled by a jump-diffusion process. However in order to estimate $\sigma(t, S_t)$ we need to obtain the paths of pure diffusion processes. To do so we rely on the method developed by Bashchenko and Marchal (2019) in order to detect and remove the jumps in the data. Then we perform the bubble detection on the synthetic pure-diffusion path and after that trade on the original path (with jumps included). The results of this classification are displayed in figure C.1 in the appendix. The performance of the trading strategy is presented in figure 6.1 where the vertical axis depicts the dollar value of the portfolio.

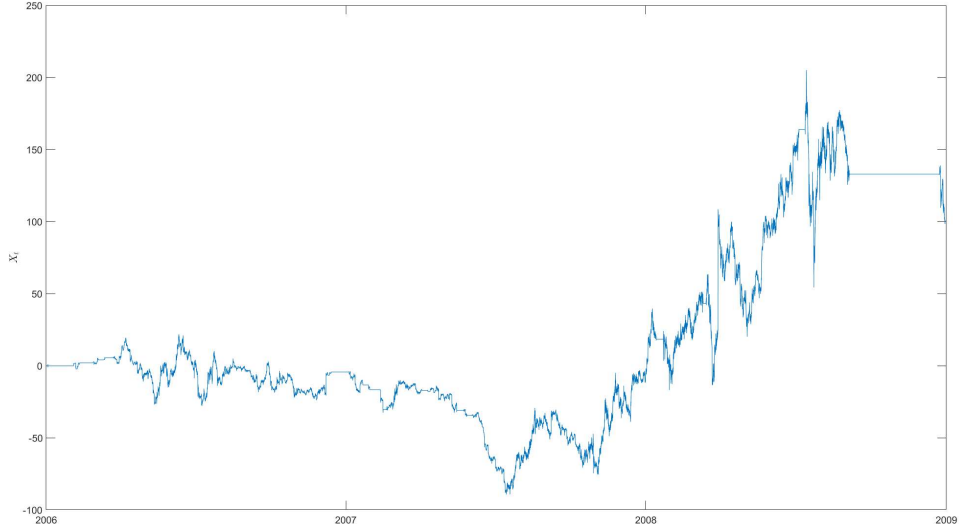

Figure 6.1: Portfolio value of our long-short trading strategy with zero net exposure.

It is worth emphasizing that our trading strategy is not simply to short high volatility stocks. Suppose that a stock (call it A) has a high volatility but is a true martingale (if $\gamma_0$ is large but $\gamma_1 \leq 1$). It is possible that a second stock (call it B) has a lower volatility than stock A but is a strict local martingale (if $\gamma_0$ is low but $\gamma_1 > 1$). Our strategy would short stock B but not A.

## 7. DISCRETE-TIME PARADIGM

As discussed in section 3, strict local martingales are a continuous-time phenomenon and do not exist in discrete-time. Thus every method (not only our neural network) detecting bubbles by relying on the theory of SLM is subject to the assumption of time continuity. We have discussed previously why we consider such assumption as not only appropriate but in fact preferred over the discrete-time paradigm. However, if the reader is a strong believer in the latter, in this section we explain what would we detect if the true price process indeed evolved in discrete-time.

In this case we assume that the stock price follows the equivalent of our continuous-time process. For this purpose, let us first time-discretize the stock price (3.6)[6] as

$$S_{t+\Delta t} = S_t + \gamma_0 S_t^{\gamma_1} \sqrt{\Delta t}\, Z \tag{7.1}$$

where the process innovations $Z \sim N(0,1)$ are iid under $\mathbb{Q}$. In discrete-time, $S_t$ is trivially a $\mathbb{Q}$-martingale for any $\boldsymbol{\gamma}$. However for $\gamma_1 > 1$ the mass of its transition density shifts to the left making it positively skewed. This implies that when $\gamma_1 > 1$, the stock price will often decrease (by a small amount) and rarely increase (but by a large amount), so that the conditional expectation is kept equal to the current value of $S_t$.

---

[6]Recall that we assumed a risk-free rate constant and equal to zero for simplicity.

In this discrete-time setting since the time horizon of the economy is finite, bubbles cannot exist. Therefore, instead of detecting asset bubbles, our network will detect periods when the stock price frequently decreases by a small amount and rarely increases by a large amount. However, since $S_t = S_t^\star$ at all times, the long-short trading strategy previously explained would not generate any profits. The fact that the strategy is profitable (as seen in figure 6.1) provides additional support for the continuous-time paradigm.

## 8. CONCLUSION

Relying on the martingale theory of asset bubbles, we show that a long short-term memory (LSTM) network is able to detect them and outperforms the current statistical estimator. On a technical level, the network performs the detection by being able to determine if a given time-series has more likely been generated by a strict local or a true martingale. The algorithm does so by estimating the functional form of the diffusion coefficient of a stochastic differential equation and identifying market regime changes. We then deploy our methodology to US equity data and show that there were multiple bubbles between 2006 and 2008. Finally, using this information we construct a zero net exposure trading strategy that shorts assets experiencing a bubble to assess the economic significance of this phenomenon.

Time continuity is a necessary assumption for the type of bubbles we detect. However this hypothesis is a source of debate among researchers. The profitability of our strategy is therefore an indirect support for the validity of using continuous-time processes to model asset prices.

# APPENDIX A

This section provides a brief intuition about neural networks and an overview of the structure of the specific network we use.

Artificial Neural Networks (ANN or simply NN) are for now one of the most powerful and widely used tool to tackle complex machine learning problems. In essence, every NN is a sequence of non-linear data transformations. A network consists of units called *neurons*, which are hierarchically organized into *layers*. Each neuron in the network is associated with its own weighting vector $W$ and bias $b$, which are the parameters that will be updated during the learning process. The neuron performs an affine data transformation (multiplying the input by its weighting vector and adding the bias), and then applies a predetermined activation function[7] to the result. All neurons of layer $l$ take as input the previous $l-1$ layer's output, and in turn pass their own output as an input for the following layer $l+1$. Neurons within one layer use the same activation function and operate independently from each other. Formally, the output of neuron number $k$ in layer $l$ is

$$a_k^{[l]} = \phi(W_k^{[l]'} x + b_k^{[l]}) \tag{A.1}$$

where $\phi$ is the activation function that is applied elementwise and $x$ is the output of all neurons of the previous layer. Passing the data through the network and obtaining the output is called *forward propagation*.

The goal of network learning is to find parameter values that will result in a minimal error between the network output and the desired output (in our case this is the labelled output). It is done iteratively. First, the input data is fed to the network and the output is obtained. Then an error function is computed, that shows how far is the network result from the target. The chosen activation functions being piecewise differentiable, a gradient descend method is used to update the parameters. This process is called *backpropagation*. Repeating forward and backpropagation allows to adjust parameters in the way that results in increasing network performance (i.e. lower error function).

An example of a simple network is the logit regression. It has two layers:[8] The first one is the input layer, with the amount of neurons being equal to the number of regressors. The second layer is the output layer with one single neuron, that has the *sigmoid* activation function $\sigma(x) = \frac{1}{1+e^{-x}}$. Parameters of this neuron are just the regression coefficients. This network has no hidden layers (that is, layers other than input and output). Networks that have one or no hidden layers are called shallow, while networks with multiple hidden layers are called *deep*. Figure A.1 provides a visualization of a deep neural network.

---

[7] Theoretically, any function can be used as an activation. However, a piecewise differentiable function is usually preferred in practice.

[8] Due to conventions, it is called a one layer network since the input layer is usually not counted.

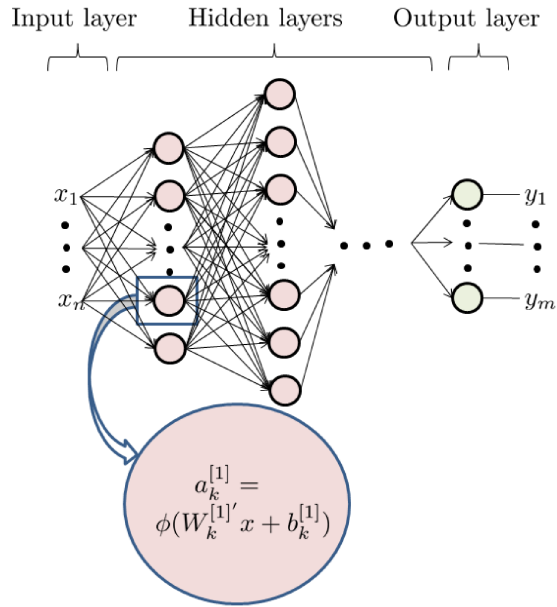

Figure A.1: Deep Neural Network

The drawback of using a general network as described above is that it is incapable of learning temporal dependencies from time-series data. To address this task a *recurrent neural network* (RNN) could be used. The recursive layer of such network has an analogue of memory, called *hidden state*. It carries the information from the previous time step and is used as additional input for the recursive layer. Formally, it processes time $t$ observation according to

$$h_t = \phi(W x_t + U h_{t-1} + b) \tag{A.2}$$

where $x_t$ is the input, $h_{t-1}$ is the hidden state from the previous time step and $W$ and $U$ are the corresponding weighting matrix for the input and hidden state respectively. $h_t$ is the updated hidden state, that is kept to treat the next observation $t + 1$ and also it serves as the output. Two major issues arise for such plain vanilla RNN. First, the memory is short-lived and not elective. There is no way to keep for a long time the important information and quickly forget the irrelevant one. The second issue, technical in nature, is the exploding/vanishing gradient.[9]

*Long short-term memory network* (LSTM) is a specific type of RNN, that mitigates both of these problems. An LSTM block has two states. One state corresponds to the working memory and is analogous to the RNN hidden state $h_t$. It is also the output of a block to the following network layer. The second one is the long-term memory mechanism, called the cell state and denoted by $c_t$. The block also has three gates (non-linear input transformation), that regulate the information flow inside:

(r) The forget/remember gate coordinates which information from the long-term memory should be kept and which one should be discarded.

(s) The input gate (or sometimes called save gate) decides which information from the input should be saved in the long-term memory.

(f) The output gate (sometimes called focus) controls the updates of the hidden state.

---

[9]Intuitively, each RNN could be unfolded into a non-recurrent network of the same length than the data series. During backpropagation, the derivative of the error function with respect to weights should be computed for every node. Due to the chain rule, it results in iterative multiplication and thus the derivative may become unstable.

Each gate is in essence just a shallow neural network itself. The parameters associated with the **r**emember, **s**ave, **f**ocus gates respectively are given by the triplet $(W_i, U_i, b_i)$ where $i \in \{r, s, f\}$.

To better grasp the intuition, let us walk along the transformation of the input $x_t$ within one LSTM block. From the previous time step the cell state $c_{t-1}$ and the hidden state $h_{t-1}$ are passed.

1. This first step is dedicated to learning which information of the existing long-term memory will be kept or forgotten. To do so, the remember gate (r) uses $x_t$ and the working memory $h_{t-1}$ to obtain the "remember" vector $r_t$ that is computed as

$$r_t = \sigma(W_r x_t + U_r h_{t-1} + b_r). \tag{A.3}$$

Here $\sigma$ defines the sigmoid activation function, that ensures values of the remember vector are between 0 (fully forget) and 1 (fully remember). $r_t$ will later be elementwise multiplied by $c_{t-1}$ to keep only the relevant information.

2. Now we decide which information should potentially be added to the long-term memory. The candidate is formed as

$$c'_t = \tanh(W_l x_t + U_l h_{t-1} + b_l) \tag{A.4}$$

where $(W_l, U_l, b_l)$ are the parameters of the **l**ong-term memory candidate formation.

3. Before it enters the long-term memory, the save gate (s) decides which part of this candidate is worth saving by computing the following quantity

$$s_t = \sigma(W_s x_t + U_s h_{t-1} + b_s). \tag{A.5}$$

As before, the activation function here is sigmoid, that ensures values between 0 and 1 and thus by elementwise multiplication allows to regulate the information flow.

4. We are ready to update the long-term memory by performing the following operation

$$c_t = c_{t-1} \otimes r_t + s_t \otimes c'_t \tag{A.6}$$

where $\otimes$ denotes an elementwise multiplication. The updated value of the long-term memory $c_t$ consists of the information remembered from the past (first term) and the newly added component (second term).

5. Finally, we can update the hidden state $h_t$. The focus gate (f) allows to concentrate on the relevant information from the long-term memory. This is done as follows

$$f_t = \sigma(W_f x_t + U_f h_{t-1} + b_f), \quad (A.7)$$
$$h_t = f_t \otimes \tanh(c_t). \quad (A.8)$$

The figure A.2 allows to represent those transformations visually.

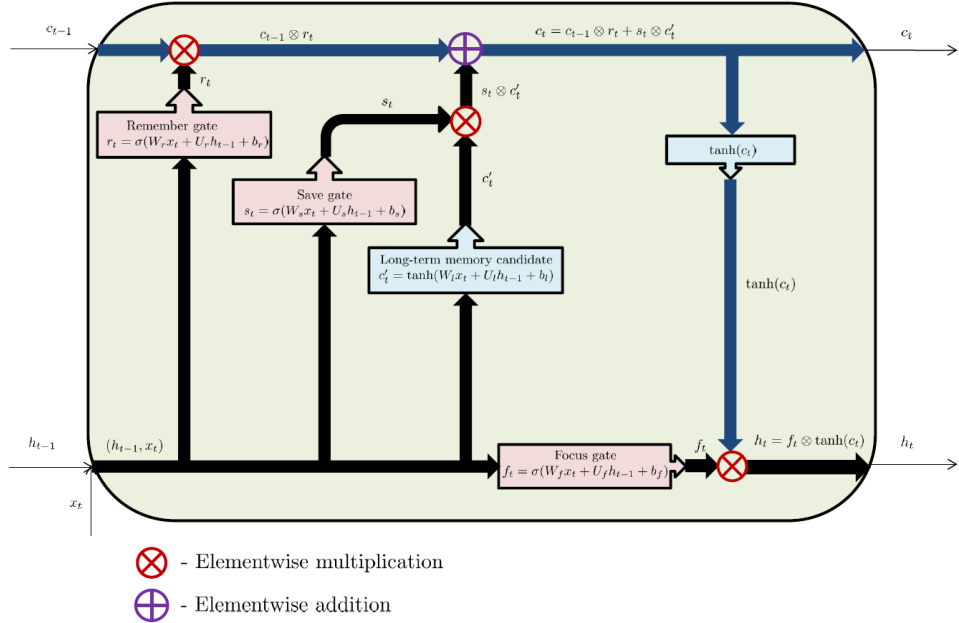

Figure A.2: Transformation of the input inside the LSTM unit

Now that we have a brief intuition about how the LSTM block works, this section concludes with a brief description of our neural network architecture. The network consists of the following layers:

1. Sequence input layer, that inputs the time-series into the network.

2. Two consecutive bidirectional LSTM layers with 100 neurons each. These layers are learning long-term dependencies from the complete sequence
    - An LSTM layer, as discussed before, allows the network to keep track of the valuable information, that was encountered long time ago, forgetting the more recent but unimportant.
    - A bidirectional layer duplicates the LSTM layer, creating two such layers one after the other. The first receives the actual time-series as an input, while the second one receives the reversed copy of the data. This allows the network to use the whole

dataset to classify points, including information that comes from the moments after.

3. Fully connected layer with two neurons, both of them being connected to all the neurons from the previous layer. Each neuron multiplies the input by the weight vector and adds the bias. This layer assembles all the features learned by the bidirectional LSTM layers to classify points into "true martingale"/"strict local martingale" categories.

4. Softmax layer with two neurons, that applies softmax function [10] to the inputs, computing the probabilities of the point belonging to one of the two categories. If the outputs of the previous layer's two neurons are $s_1$ and $s_2$, the neurons of this layer will compute $p_i = \frac{e^{s_i}}{\sum_{j=1}^{2} e^{s_j}}$ for $i = 1, 2$.

5. Classification output layer, that computes the cross-entropy[11] for the classification problem for multiple non-intersecting classes in order to construct the error function (that will be minimized).

A schematic representation of a standard classification network with one biLSTM layer can be found on the figure A.3.

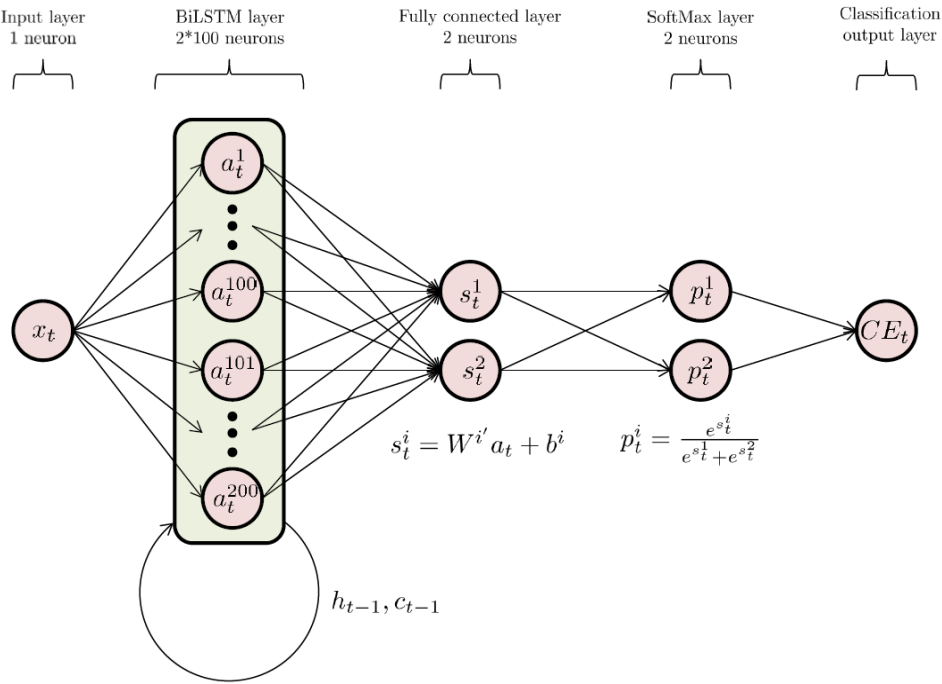

Figure A.3: Standard LSTM network for classification.

---

[10] The softmax function (also called normalized exponential) takes as input a real vector and transforms it into a probability distribution. Formally, for a vector $v \in \mathbb{R}^n$, the softmax function $g : \mathbb{R}^n \to \mathbb{R}^n$ is defined as $g(v)_i \triangleq \frac{e^{v_i}}{\sum_{j=1}^{n} e^{v_j}}$ for $i = 1, ..., n$.

[11] The cross-entropy of two probability distributions $\mathbb{P}$ and $\mathbb{P}^*$ is defined as $H(\mathbb{P}, \mathbb{P}^*) = E^{\mathbb{P}}\left[-\log \mathbb{P}^*\right]$.

## APPENDIX B: DISCUSSION ON LOCAL MARTINGALES

In this section we aim at building intuition about true and strict local martingales. We will do so through the famous example of the doubling strategy. Imagine that a gambler is betting on the outcome of a coin toss. If the coin comes out head, he wins his bet. If it comes out tail he looses his bet.

Now assume that he chooses the following strategy: if he looses round $n$, he doubles his previous bet for the next round. He does so until he wins. As soon as he wins for the first time, he takes his gain and stops playing. Clearly, if the player is allowed to take infinite credit, he can bet endlessly until the coin comes out head and thus is guaranteed to walk away with a net gain of 1 dollar. However, for every finite number of trials it is a fair game: with high probability the player wins 1, and with very small probability he loses all his previous bets, such that the expected net gain is 0. Therefore his personal wealth is a true martingale.

However if the gambler is allowed to bet infinitely fast in a finite time interval, his wealth process turns into a strict local martingale and is expected to increase.

### B.1. BETTING IN DISCRETE-TIME

Let $X_t^\pi$ denote the dollar value at time $t$ of the portfolio that invests in the strategy $\pi$ described above. We set $X_0^\pi = 0$ such that the gambler starts with zero initial wealth. Let $Z$ be the random variable that represents the outcome of the coin toss. It takes the value 1 if the coin lands on head and -1 otherwise. The coin is unbiased such that $\mathbb{P}(Z_n = 1) = \mathbb{P}(Z_n = -1) = 1/2$. It follows that

$$X_n^\pi = \begin{cases} 1 & \text{if } X_{n-1}^\pi = 1, \\ X_{n-1}^\pi + (1 - X_{n-1}^\pi) Z_n & \text{otherwise.} \end{cases}$$

As we see, until the first time the coin comes out heads, the value of the strategy $X_n^\pi$ is negative and the debt grows exponentially. As soon as the coin lands on head, the winning bet covers the previously occurred losses and provides the net gain of 1 dollar. It is easy to see that this is a martingale.

For every finite number of tosses $N$ the probability of loss (i.e the probability of coin landing only on the tails all N times) is $\frac{1}{2^N}$ and the corresponding loss in this case is the sum of all the lost bets, including the current one, so it is $(-2^N + 1)$. The probability of winning is then $1 - \frac{1}{2^N}$ and the net gain in case of victory is 1.

Even though the expected gain stays zero in discrete-time independently of $N$, it is insightful to analyze what happens to the distribution as the number of bets rises to understand the continuous-time framework in the next section. For that, we simulate $10^7$ sample paths of such game, each representing one different gambler. We present the histograms of the players' wealth after $n$ rounds below. In line with the reasoning above, after the first coin toss in half of the cases the player leaves as a winner with \$1 and in the other half the player looses \$1 as displayed in figure B.1. Increasing the number of tosses to 4, we see in figure B.2 a shift in the distribution: most of the mass is concentrated on the winners side, with approximately $\frac{15}{16}$ of players being the winners of \$1 and $\frac{1}{16}$ of gamblers loosing \$15. This phenomenon accentuates as we increase further *n*: the probability of loosing decreases but the potential loss increases, thus decreasing the skewness of the distribution. The loss probability will disappear when we work in continuous-time, transforming the gambler's wealth into a strict local martingale.

While in theory the expectation of this discrete-time process is always 0, when increasing the number of trials (for example up to 25) we run into the *finite-sample problem*. The probability of loss becomes so small that our number of samples is just not high enough: simulations will tell us that among $10^7$ players after 25 rounds there was no losers and everyone eventually walked away with $1. In this case the average terminal wealth is clearly 1. Here we already approximate a strict local martingale behavior that should only exist in continuous-time. But this picture is misleading and arises solely due to the computational limitations. We should remember that this is still a fair game! The probability of the loss is, however small, still nonzero, and the potential loss is so huge that it balances the mass of winners in expectation.

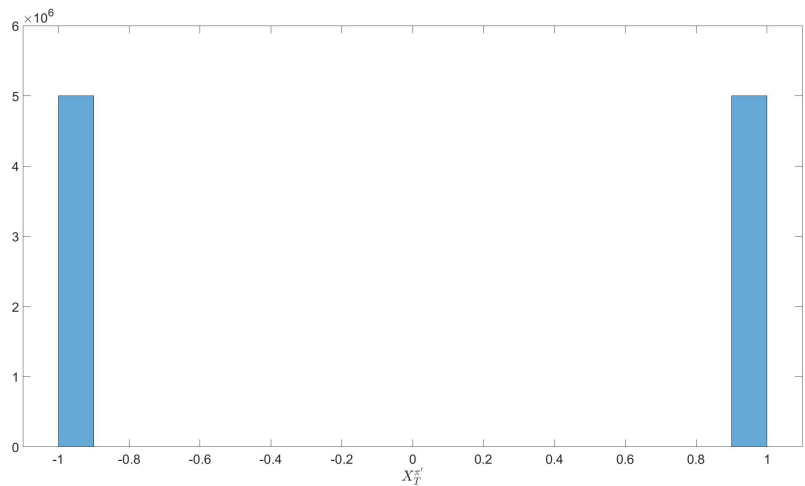

Figure B.1: Distribution of the terminal wealth of the gamblers after 1 trial.

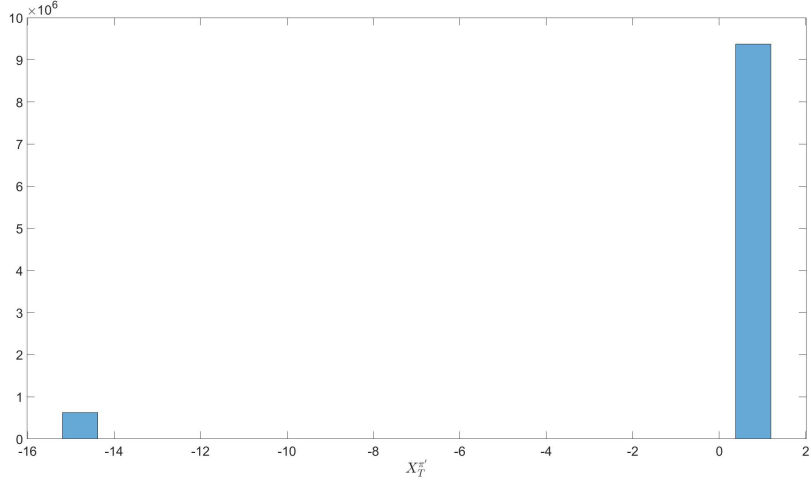

Figure B.2: Distribution of the terminal wealth of the gamblers after 4 trials.

### B.2. BETTING IN CONTINUOUS-TIME

The previously described strategy is pathological in the sense that even though for every finite number of trial it is fair, it stops being fair as soon as we allow for infinitely many trials. This pathology is translated into the strict local martingale property of the continuous-time process. Imagine that the player can speed up the time, such that after each lost trial he can bet faster and faster. In such a fantastic world the player can place infinitely many bets (thus eventually win \$1) in a finite time interval (say 1 hour). Mathematically, we construct the following continuous-time process:

$$Y_t = \begin{cases} X_n^\pi & \text{if } 1 - \frac{1}{n} \le t < 1 - \frac{1}{n+1}, \\ 1 & t \ge 1. \end{cases}$$

This process represents the dollar value of a portfolio which employs this doubling strategy. As a result, the player always ends up with the net gain of 1. This process is clearly not a martingale, since $\mathbb{E}[Y_0] = 0 \ne \mathbb{E}[Y_1] = 1$. However, it is a local martingale (under the localizing sequence of stopping times chosen as, for example, $\tau_n = \inf\{t : |X_t| \ge n\}$).

It is important to point that not all strict local martingales arise due to a doubling strategy. As pointed out in Dumas and Luciano (2017), in this example the strict local martingale arises due to the behavior of the agent (which bets faster and faster) while the underlying stochastic process $Z$ (the coin) itself has always the same distribution after each round since the variable is iid. In the bubble detection part of this paper, the process (3.6) itself might be a strict local martingale irrespective of the betting behavior of the investor. Therefore it is important to understand that when we detect strict local martingales (i.e. bubbles) and implement a long-short trading strategy, we do not rely on a doubling mechanism.

## B.3. APPROXIMATION OF STRICT LOCAL MARTINGALES BY DISCRETE-TIME PROCESSES

The final point of our discussion concerns the simulation of strict local martingales, a phenomenon that exists solely in continuous-time. Since computers cannot simulate continuous variables, we have to rely on a discretization scheme. So in fact, we are left with a discrete-time process, trying to study a phenomenon, that exists purely in continuous-time. However, this appears to cause no complication for our purpose. As we have seen in subsection B.1, the probability mass of the gambler's wealth is sparsely distributed. As $n$ rises, a higher probability is concentrated on the winning point, while a very small and further decreasing probability is shifting to the left, corresponding to loosing higher and higher amounts. At some point the probability becomes so small, that among the whole simulation set there is not enough samples for this probability to realise and create at least one loosing path. As a consequence, we end up with a process $X^\pi$ that approximates a strict local martingale behavior. Even though we still work in discrete-time and theoretically this process should be a martingale (staying at 0 in expectation), due to the absence of at least one loosing path the simulated process in fact increases on average.

The same logic applies to our simulations of the stock price $S_t$. We use Monte-Carlo to generate paths of the discretized version of the price in equation (7.1) where $\gamma_1 > 1$. Its continuous-time counterpart is a strict local martingale which is a supermartingale (since is bounded below by 0) and so has to go down in expectation. However even the simulated discrete-time process decreases on average while in theory it should not. This happens exactly because of an extremely sparse probability distribution as discussed in subsection B.1. With very high probability the process goes down (resembling an SLM) while with very small probability it goes extremely high and thus balances the average to stay a true martingale. Since we simulate a finite number of paths, this highly unlikely exploding path just does not appear, leaving us with paths that resemble the SLM.

If we choose a time interval $\Delta t$ of $10^{-3}$ with maturity $T = 5$ for the simulations, we observe a decreasing average even if we simulate over $10^8$ different paths. Decreasing $\Delta t$ further for the same $T$ will accentuate this phenomenon.

# APPENDIX C: GRAPHS

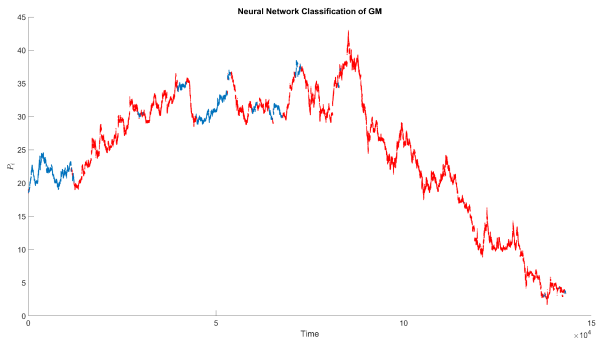

(a) GM

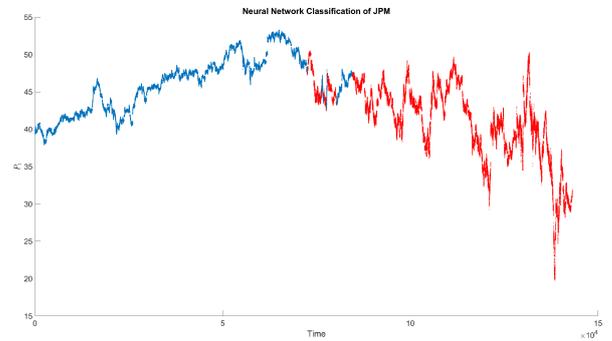

(b) JPM

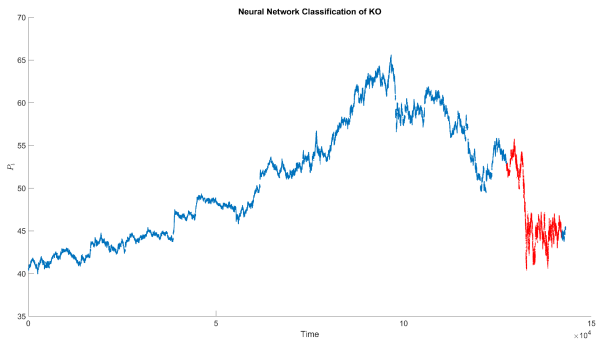

(c) KO

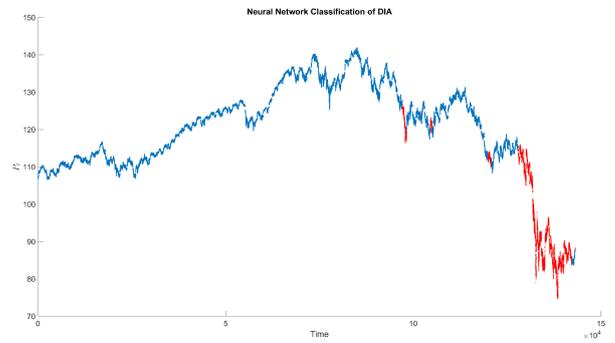

(d) DIA

Figure C.1: Bubbles detection from 2006 to 2008 included (3 years) using our LSTM network. The blue line represents the price under a non bubbly regime. The red line represents periods when the asset is in a bubble regime.

The figures for the remaining stocks in the Dow Jones can be found on the personal webpages of the authors.